\documentclass[letterpaper,12pt]{report}
\usepackage[T1]{fontenc} 
\usepackage[latin1]{inputenc} 
\usepackage[dvips]{graphicx}
\usepackage[left=3cm, right=3cm, top=3cm, bottom=3cm]{geometry}
\usepackage{longtable}
\usepackage{amsthm}
\usepackage{amsfonts}
\usepackage{amsmath}
\usepackage{amssymb}
\usepackage{color} 
\title{\LARGE \bf Josephson Effect in Superconductors and Superfluids}
\author{\bf Anna Posazhennikova \\
{\it Department of Physics, University of Konstanz,D-78457 Konstanz, Germany}}
\begin{document}
\maketitle
\tableofcontents
\listoffigures

\chapter{Macroscopic phase coherence and Josephson effect}

In this Chapter we consider systems, that at first glance may seem very
different: (i) superconductors, which at temperatures larger than the critical
temperature $T_c$ ($T_c$ in conventional superconductors is of order of a few
 $K$) turn metallic and are described by
the Fermi Liquid theory, (ii) superfluid liquid He: bosonic $^4$He, $T_c=2.17K$ and fermionic
$^3$He, $T_c=0.0025K$; and (iii) Bose-Einstein condensates of cold
alkali atoms (with $T_c$ of the order of nano Kelvin).  
All these systems have one important property in common - at low temperatures
they possess {\it macroscopic coherence}.

In previous Chapters it was discussed that a superconducting state is a
state with a broken $U(1)$--symmetry and is thus characterized by a complex
order parameter 
\begin{equation}
\Psi=\Psi({\bf r})=|\Psi({\bf r})|e^{i\phi({\bf r})}
\label{wf}
\end{equation}
which represents a macroscopic wave function of a superconductor. The presence of the order
parameter means that at any given moment the phase difference of
$\Psi$-functions between any two macroscopically separated points in the
superconductor is fixed, so that the whole sample acquires macroscopic phase
coherence, or in other words, long-range order develops. 
One can describe a condensate of cold bosons or a superfluid He system in a
similar way, since these systems are also phase coherent.

Phase coherence leads to a number of specific {\it quantum} effects. For example, in
superconductors it causes the quantization of  magnetic flux first considered by
London. One of the most celebrated manifestations of the phase coherence property
is however the Josephson effect which is the subject of the current Chapter.

\chapter{Josephson Effect in superconductors}
\label{sc}

In 1962 Brian Josephson predicted a curious effect occurring in a system of  two weakly-linked
superconductors \cite{Josephson}. He demonstrated that a direct current can flow between two
superconductors coupled via an insulating thin layer although no external
voltage is applied. Furthermore, he showed that an external voltage would give
rise to a rapidly oscillating current. Josephson's theory of a current induced
between two superconductors was rather fast confirmed experimentally \cite{Barone}. 

\begin{figure}[!htb]
\begin{center}
\includegraphics[scale=0.5,angle=0.1,keepaspectratio]{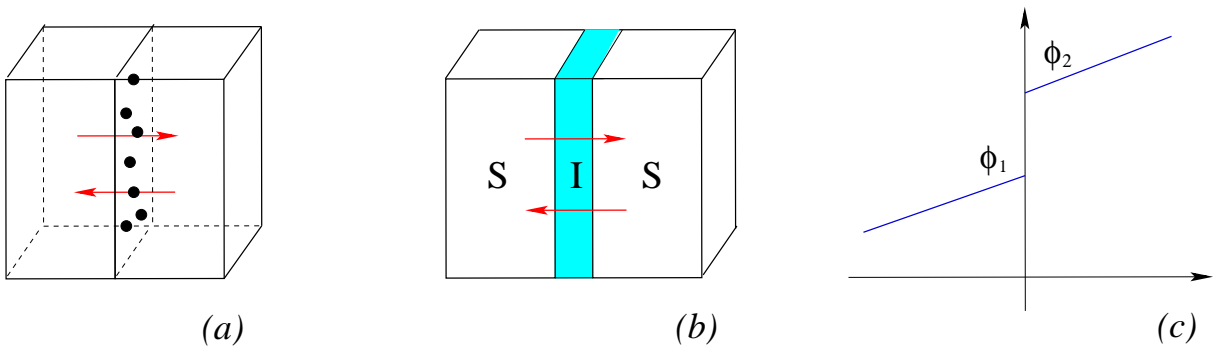}
\end{center}
\caption{\em Superconductor with an ``impurity plane'' inside
  (a). Supercurrent will persist even if the plane is replaced by a thin
  insulating layer (b). The phase jump at the barrier separating two
  superconductors (c).}
\label{fig1}
\end{figure}

The Josephson effect can be understood by the following simple considerations.  First of
all it is important to observe that 
 a gradient of the phase gives
rise to a current
\begin{equation}
j=\frac{N_se\hbar}{2m}\nabla \phi,
\label{current_grad}
\end{equation}
where $N_s$ is the number of superconducting electrons. Imagine now that
instead of a uniform superconductor we deal with a superconductor with
impurities: point-like, randomly distributed non-magnetic impurities. As we
know, for an s-wave superconductor impurities do not affect the critical
temperature, they do not destroy phase coherence in a superconductor (in
accordance with Anderson's theorem \cite{Anderson59}). 
As a consequence supercurrents \eqref{current_grad} can flow through the system without
a problem. This situation does not change if instead of impurities distributed
in the bulk we have impurities distributed solely in a plane inside a
superconductor (Fig.\ref{fig1}(a)).

Supercurrents proportional to the the fixed phase gradient can flow through the
plane, even though the mean spacing between the impurities is smaller than the
coherence length of a superconductor. This intuitively clear picture can be
generalized to the case of a somewhat more complicated system: two
superconductors separated by a thin insulating layer. This nontrivial
generalization was realized by Josephson and led him to the prediction of two
effects named after him: a.c. and d.c. Josephson effects.  

These Josephson effects are pure quantum phenomena, because electrons travel from one superconductor to the other by means
of quantum mechanical tunneling through the barrier separating  the two
superconducting systems. The presence of a barrier, or inhomogeneity leads
to the fact that the phase has a jump at the barrier (Fig. \ref{fig1}(c)). The
supercurrent through the barrier is then driven by the phase
difference $\phi=\phi_2-\phi_1$. Thus, a supercurrent can flow between
two superconductors provided they are separated by a sufficiently thin
insulating layer. This effect is referred to as {\it first} or  stationary, or
{\it d.c. Josephson effect}. In this case the potential difference through
the barrier is equal to zero. Note, that while the superconducting coherence
length is of the order of $\xi \sim 10^4 $~\AA, the thickness of the insulting
layer should be of the order of $10-20$~\AA , i.e. thousands of times smaller
than $\xi$.

When a finite voltage bias $V$ is applied to an SIS junction, the {\it second},
non-stationary or {\it a.c. Josephson effect} can be observed. In this case
the current will
be oscillating between the two superconductors with a frequency $\omega$ proportional
to the applied bias
\begin{equation}
\hbar \omega=2e V.
\label{freq}
\end{equation} 

Since in a superconductor electrons are bound into Cooper pairs, these 
 pairs participate in the tunneling  across a barrier between two
superconductors. The energy $2eV$ is then just the difference in the energy of a
Cooper pair in passing from one superconductor to the other. 

One can estimate a Josephson current from simple electrodynamic
considerations \cite{Abrikosov}. 
First of all we observe, that the
appearance of a current in the system is related to the excess energy $\Delta E=E(\phi)-E(\phi=0)$
associated with the junction between the two superconductors. It is clear that
$\Delta E$ should be proportional to the product of two superconducting gaps 
$\Delta_1\Delta_2$, because if one of superconductors is absent, the excess
energy vanishes. Apart from that the excess energy should be real, 
one can therefore conjecture its simplest possible form as follows
\begin{equation}
\Delta E=\alpha\int dx dy \left(
  |\Delta_1\Delta_2|-\frac{1}{2}(\Delta_1\Delta_2^*+c.c.)\right)=2\alpha\int
dx dy|\Delta_1\Delta_2|(1-\cos \phi),
\label{int_energy}
\end{equation}
where $\alpha$ is a phenomenological constant describing the
coupling between the two superconductors in the  $x-y$ junction plane. 

One can derive now from the gauge invariance principle that the supercurrent is
proportional to $(\sin \phi)$. The gauge invariance requires the replacement
\begin{equation} 
\nabla \phi \rightarrow \nabla \phi-\frac{2e}{\hbar c}{\bf A},
\label{gauge}
\end{equation}
where ${\bf A}$ is the vector potential. For simplicity we can chose the
vector potential perpendicular to the $x-y$ plane, so that integration along
that axes of the r.h.s. of \eqref{gauge} gives
\begin{equation}
\phi_1-\phi_2-\frac{2e}{\hbar c}\int_L^Rdz A_z,
\end{equation}
where $\phi_1$ is the phase of the ``left'' superconductor, and $\phi_2$ is
the phase of the ``right'' superconductor. 
The excess energy becomes
\begin{equation}
\Delta E=2\alpha\int
dx\, dy|\Delta_1\Delta_2|\left[1-\cos\left(\phi_1-\phi_2-\frac{2e}{\hbar c}\int_L^Rdz A_z \right)\right].
\end{equation}

Variation of this energy with respect to the potential gives
\begin{equation}
\delta(\Delta E)=-\frac{2e}{\hbar c}\alpha \int dx\, dy\, dz\, |\Delta_1\Delta_2|
\sin\left(\phi_1-\phi_2-\frac{2e}{\hbar c}\int_L^Rdz A_z \right)\Big]\delta A_z.
\end{equation}

From electrodynamics we know that
\begin{equation}
\delta E=-\frac{1}{c}\int {\bf j}\;\delta{\bf A}\; d V,
\end{equation}
and  we get for the supercurrent
\begin{equation}
j=\frac{2e}{\hbar}\alpha|\Delta_1\Delta_2|\sin\left(\phi_1-\phi_2-\frac{2e}{\hbar c}\int_L^Rdz A_z \right).
\end{equation}
For zero vector potential we obtain the famous expression for the Josephson current
\begin{equation}
j=j_c\sin(\phi_1-\phi_2).
\label{jos_cur}
\end{equation}
The current vanishes for $\phi_1-\phi_2=0$. The so-called critical current $j_c$
should be calculated microscopically (see Section \ref{ms}).

The basic result \eqref{jos_cur} can be obtained in a different way. 
The following derivation in terms of a two-level system is due to Feynman
\cite{Feynman}. As we discussed already, each superconductor can be considered as a macroscopic quantum
state described by a wave-function $\Psi_{1,2}$ \eqref{wf}.
Since the coupling between the
superconductors is very weak, the state vector describing the coupled system
can be written in a simple form
\begin{equation}
|\Psi \rangle=\Psi_1 |1 \rangle +\Psi_2 |2 \rangle.
\label{state}
\end{equation}
The density of superconducting electrons in the left (right) superconductor,
described by the state $|1 \rangle$ ($| 2 \rangle$) is defined as
\begin{equation}
n_i=|\Psi_i|^2=\langle i| \Psi \rangle \langle \Psi| i \rangle,
\label{dens}
\end{equation}
$i=1,2$. 
The Schr\"odinger equation of motion for the state vector \eqref{state} reads
\begin{equation}
i\hbar \partial_t |\Psi \rangle=H|\Psi \rangle,
\label{schr}
\end{equation}
with the Hamiltonian 
\begin{equation}
H=H_1+H_2+H_{int}.
\end{equation}
Here $H_i=E_i|i\rangle \langle i|$, ($i=1,2$) and the coupling between the superconductors can be
written in analogy with \eqref{int_energy}
\begin{equation}
H_{int}=-\frac{\alpha}{2}(|1\rangle \langle 2|+|2\rangle \langle 1|). 
\end{equation}
Projection of the Eq. \eqref{schr} on the two states gives the equations of
motion for two weakly coupled superconductors
\begin{eqnarray}
i\hbar \partial_t\Psi_1=E_1\Psi_1 -\frac{\alpha}{2}\Psi_2, \nonumber \\
i\hbar \partial_t\Psi_2=E_2\Psi_2 -\frac{\alpha}{2}\Psi_1.
\end{eqnarray}
Remembering that $\Psi_i$ can be expressed in terms of superconducting
densities (see \eqref{dens})
\begin{equation}
\Psi_i=\sqrt{n_i}e^{i\phi_i},
\end{equation} 
we can derive the final equations in terms of the densities and phases
\begin{eqnarray}
\partial_t n_1&=&\frac{\alpha}{\hbar}\sqrt{n_1 n_2} \sin(\phi_1-\phi_2)=-\partial_t
n_2, \\
\partial_t (\phi_2-\phi_1)&=&\frac{1}{\hbar}(E_1-E_2)+\frac{\alpha}{2\hbar}\frac{n_1-n_2}{\sqrt{n_1n_2}}\cos(\phi_1-\phi_2).
\label{phase}
\end{eqnarray}
For equal densities $n_1=n_2\equiv n$ we get
\begin{equation}
\partial_t (\phi_2-\phi_1)=\frac{1}{\hbar}(E_1-E_2).
\end{equation}
The pair current density is given by
\begin{equation}
j\equiv \partial n_1=-\partial n_2=j_c \sin(\phi_1-\phi_2).
\label{eq1}
\end{equation}
with $j_c=\alpha n/\hbar$ for equal densities. One should
note, that the densities $n_1$ and $n_2$ are considered to be constant (we
will see that this is not the case in a Bose Josephson junction), their time
derivative is however not constant due to the presence of the external current source
which continuously replaces the pairs tunneling across the barrier. 

The
presence of the potential difference $V$ is easily taken into account in our
equations. In two isolated superconductors the energy terms are given by the
chemical potentials $E_i=2 \mu_i$ ($i=1,2$). A d.c. potential difference will
shift these chemical potentials by $eV$, so that $E_1-E_2=2eV$, and
Eq. \eqref{phase} becomes
\begin{equation}
\partial_t (\phi_2-\phi_1)=\frac{2eV}{\hbar}.
\label{eq2}
\end{equation}
The two equations \eqref{eq1} and \eqref{eq2} constitute two main relations of
the Josephson effect, which we discussed at the beginning of this Section. For $V=0$ the phase difference is constant, so that a
finite current density with a maximum value $j_c$ can flow through the barrier
with zero voltage drop across the junction. This is the essence of the d.c. Josephson effect. With a finite potential difference $V$ applied to
the junction there appears an alternating current
\begin{equation}
j=j_c \sin(\phi_0+\frac{2e}{\hbar}Vt)
\end{equation}
with a frequency \eqref{freq}. This corresponds to the a.c. Josephson effect. 

In the following we derive the microscopic expression for the critical
Josephson current.

\chapter{Microscopic derivation of a critical superconducting current}
\label{ms}

The microscopic approach for the calculation of critical current was suggested
by Anderson \cite{Anderson}, and Ambegaokar and Baratoff \cite{Ambegaokar}. Their method is based on the
so-called {\it tunneling Hamiltonian}. In this approach the details of the
interface are not taken into account and instead two weakly coupled superconductors
described by Hamiltonians $H_1$ and $H_2$ in the absence of tunneling are
considered, whose coupling in the first order perturbation theory is described by the
tunneling term in the Hamiltonian $H_T$
\begin{equation}
H=H_1+H_2+H_T.
\label{ham}
\end{equation} 
$H_T$ has a simple form
\begin{equation}
H_T=\sum_{{\bf pq}\sigma}T_{{\bf pq}}a^{\dagger}_{{\bf p}\sigma}b_{{\bf
    q}\sigma}+T_{{\bf pq}}^*b^{\dagger}_{{\bf q}\sigma}a_{{\bf p}\sigma},
\label{ham_tun}
\end{equation}
where $a$ are the fermionic operators of the ``left'' superconductor, and $b$
are the fermionic operators of the ``right'' superconductor, $\sigma$ is a
spin index, ${\bf p}$ and ${\bf q}$ are the momenta of electrons. Due to the
time-reversal invariance of the Schr\"odinger equation ($t\rightarrow -t$,
$\Psi\rightarrow \Psi^*$)  the matrix elements $T_{{\bf pq}}$ have the
property $T_{{\bf pq}}=T^*_{{\bf -p,-q}}$.
The Hamiltonian \eqref{ham_tun} conserves the number of particles in the
system $N_1+N_2$, where
\begin{equation}
N_1=\sum_{{\bf p}\sigma}a^{\dagger}_{{\bf p}\sigma}a_{{\bf p}\sigma}, \quad N_2=\sum_{{\bf p}\sigma}b^{\dagger}_{{\bf p}\sigma}b_{{\bf p}\sigma}.
\end{equation} 

The current is related to the change of the number of
particles with time and is therefore by definition
\begin{equation}
I=e\langle \dot N_1 \rangle=-e\langle \dot N_2 \rangle.
\label{curr_def}
\end{equation}
The equation of motion for the operator $N_i$ reads
\begin{equation}
i\hbar\dot N_i=[N_i,H_T],
\label{eom_number}
\end{equation}
where we took into account that the operator $N_i$ commutes with $H_1$ and
$H_2$ ($i=1,2$). With \eqref{eom_number} the expression for the
current \eqref{curr_def} becomes
\begin{equation}
I=-\frac{ie}{\hbar}\left(\sum_{{\bf pq}\sigma}T_{{\bf pq}}\langle a^{\dagger}_{{\bf p}\sigma}b_{{\bf
    q}\sigma}\rangle-T_{{\bf pq}}^*\langle b^{\dagger}_{{\bf q}\sigma}a_{{\bf
    p}\sigma}\rangle  \right).
\label{curr_av}
\end{equation}
One can proceed with the derivation of the Josephson current in several ways, see for instance
\cite{Abrikosov,Kulik,Mahan}. 
Here we suggest a rather straight-forward derivation based on the nonequilibrium
Keldysh technique \cite{Keldysh,Rammer}. We apply then our general results to
a problem of a stationary Josephson current between two superconductors. 

\begin{figure}[!htb]
\begin{center}
\includegraphics[scale=0.35,angle=0.1,keepaspectratio]{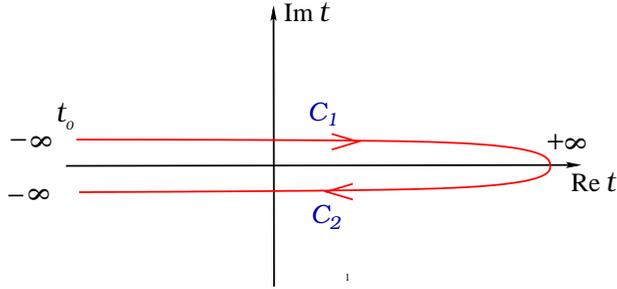}
\end{center}
\caption{\em The Keldysh path on the time-plane along which the nonequilibrium
Green's function is defined. }
\label{keld}
\end{figure}

One can introduce the so-called Keldysh Green's function \cite{Keldysh} which is defined in
the following way
\begin{equation}
G^K_{{\bf qp}\sigma}(t,t')=2i\langle a^{\dagger}_{{\bf p}\sigma}(t')b_{{\bf
    q}\sigma}(t) \rangle.
\label{func_keld}
\end{equation}
The Keldysh Green's function \eqref{func_keld} is a part of a general
path-ordered Green's
function 
\begin{eqnarray}
G_{{\bf pq}\sigma}(t,t')=-i \langle
T_{C}\; b_{{\bf q}\sigma}(t)a^{\dagger}_{{\bf
    p}\sigma}(t')   \rangle,
\label{gen_keld}
\end{eqnarray} 
which is defined on a so-called Keldysh contour $C=C_1+C_2$ showed in
Fig. \ref{keld}, $T_C$ is a time-ordering operator along this contour. 
It is convenient to separate the Keldysh contour on the upper ($C_1$) and
lower ($C_2$) contours and to  present the function \eqref{gen_keld} in a matrix form
\begin{equation}
\hat G=\left(\begin{array}{cc} G_{11} & G_{12} \\
G_{21} & G_{22} 
\end{array}
\right),
\label{matrix_keld}
\end{equation}
where indices $1$ and $2$ refer to the upper or lower Keldysh contour
respectively. By applying the usual rotation operator
\begin{equation}
R=\frac{1}{\sqrt{2}}\left(\begin{array}{cc} 1& 1 \\ -1& 1 \end{array}\right)
\label{rot}
\end{equation}
to the matrix \eqref{matrix_keld}: $\hat G\rightarrow R^{-1}\hat G R$ we get
the nonequilibrium Green's function in the Larkin-Ovchinnikov representation \cite{Larkin75}
\begin{equation}
\hat G=\left(\begin{array}{cc} G^R & G^K \\
0 & G^A 
\end{array}
\right).
\end{equation}
The relation between $G_{11}$, $G_{12}$, $G_{21}$, $G_{22}$ and retarded
$G^R$, advance $G^A$ and Keldysh $G^K$ Green's functions can be trivially
derived from \eqref{matrix_keld} and \eqref{rot}. For details see \cite{Keldysh,Rammer}.

The Keldysh Green's function \eqref{func_keld}
is useful, because we can immediately express the current \eqref{curr_av} in terms of such functions
\begin{equation}
I=-\frac{e}{2\hbar}\sum_{{\bf pq}\sigma}\left(T_{{\bf pq}}G^K_{{\bf
      qp}\sigma}(t)-T_{{\bf pq}}^*G^K_{{\bf pq}\sigma}(t)  \right).
\end{equation}
One can readily see that $[G^K_{{\bf pq}\sigma}(t)]^*=-G^K_{{\bf qp}\sigma}(t)
$, so that the expression for the current becomes even simpler
\begin{equation}
I=-\frac{e}{\hbar}Re\left[\sum_{{\bf pq}\sigma}T_{{\bf pq}}G^K_{{\bf
      qp}\sigma}(t) \right].
\end{equation}

We need thus to calculate the Keldysh Green's function $G^K_{{\bf
    qp}\sigma}(t)$. For simplicity we proceed with our calculations in the
first order of perturbation theory. We also consider only $G_{{\bf
    pq}\uparrow}$ function, because 
$G_{{\bf pq}\downarrow}$  can be  derived analogously.

The Green's function of the system described by the Hamiltonian \eqref{ham} in the first
order of  perturbation theory reads
\begin{eqnarray}
G^{(1)}_{{\bf pq}\uparrow}(t,t')=-i \langle
T_{C}\left(1-i\int_{C}d\tau H_T(\tau)\right)b_{{\bf q}\uparrow}(t)a^{\dagger}_{{\bf
    p}\uparrow}(t')   \rangle.
\label{GF1}
\end{eqnarray} 
 According to the BCS theory only electrons with opposite
momenta and spins are allowed to pair, hence we get the expression
\begin{equation}
G^{(1)}_{{\bf pq}\uparrow}(t,t')=T_{\bf -p,-q}\langle  T_{C}\int_{C}d\tau
b_{{\bf q}\uparrow}(t)
b_{-{\bf q}\downarrow}(\tau) 
a^{\dagger}_{-{\bf p}\downarrow}(\tau) a^{\dagger}_{{\bf p}\uparrow}(t') \rangle.
\label{GF3}
\end{equation}
We assume, that in the simplest approximation the critical supercurrent
is carried by Cooper pairs, and by applying Wick's theorem \cite{AGD} to \eqref{GF3} we get 
\begin{equation}
G^{(1)}_{{\bf pq}\uparrow}(t,t')=-T_{\bf -p,-q}\int_{C}
d\tau \, \mathcal{F}_{\uparrow\downarrow}({\bf
  q},t-\tau)\overline{\mathcal{F}}_{\downarrow\uparrow}({\bf p},\tau-t').
\label{matrix_GF}
\end{equation}
Here we  introduced anomalous Gor'kov functions for a superconductor \cite{AGD}
\begin{eqnarray}
\mathcal{F}_{\uparrow\downarrow}({\bf q},t-t')&=&-i\langle T_{C}\, b_{{\bf
    q}\uparrow}(t)b_{-{\bf q}\downarrow}(t') \rangle
\nonumber \\
\overline{\mathcal{F}}_{\downarrow\uparrow}({\bf p},t-t')&=&-i\langle T_{C}\, a_{-{\bf
    p}\downarrow}^{\dagger}(t)a_{{\bf p}\uparrow}^{\dagger}(t') \rangle.
\end{eqnarray}
These Green's function do not depend on the sign of momentum ${\bf
  p}$, but the order of spin indices does matter (replacing
$\uparrow\downarrow$ with $\downarrow\uparrow$ will give an extra minus sign).

In a lengthy but straightforward calculation \cite{Rammer} one can extract
the Keldysh part of the matrix Green's function \eqref{matrix_GF}
\begin{equation} 
[G^{(1)}_{{\bf
    pq}\uparrow}(t,t')]^K=-T_{pq}^{*}\int_{-\infty}^{\infty}d\tau[\mathcal{F}_{\uparrow\downarrow}^{R}({\bf q},t-\tau)\overline{\mathcal{F}}^{K}_{\downarrow\uparrow}({\bf p},\tau-t')+\mathcal{F}_{\uparrow\downarrow}^{K}({\bf q},t-\tau)\overline{\mathcal{F}}^{A}_{\downarrow\uparrow}({\bf p},\tau-t')],
\end{equation}
 The Fourier transformation
of this expression gives
\begin{equation}
[G^{(1)}_{{\bf
    pq}\uparrow}(\omega)]^K=-T_{pq}^{*}[\mathcal{F}_{\uparrow\downarrow}^{R}({\bf q},\omega)\overline{\mathcal{F}}^{K}_{\downarrow\uparrow}({\bf p},\omega)+\mathcal{F}_{\uparrow\downarrow}^{K}({\bf q},\omega)\overline{\mathcal{F}}^{A}_{\downarrow\uparrow}({\bf p},\omega)].
\end{equation}

The current then becomes
\begin{equation}
I=\frac{2e}{\hbar}Re\left[\sum_{{\bf pq}}|T_{{\bf pq}}|^2\int_{0}^{\infty}\frac{d\omega}{2\pi} \left(\mathcal{F}_{\uparrow\downarrow}^{R}({\bf q},\omega)\overline{\mathcal{F}}^{K}_{\downarrow\uparrow}({\bf p},\omega)+\mathcal{F}_{\uparrow\downarrow}^{K}({\bf q},\omega)\overline{\mathcal{F}}^{A}_{\downarrow\uparrow}({\bf p},\omega)\right)  \right],
\end{equation}
the factor of ``2'' appears because the contribution from $G^K_{{\bf
    pq}\downarrow}$ is equivalent to the contribution from $G^K_{{\bf
    pq}\uparrow}$.

This expression simplifies greatly in equilibrium, in which case the Keldysh
Green's function can be expressed as
\begin{equation}
G^K(\omega)=(G^R(\omega)-G^A(\omega))(1-2f(\omega)),
\end{equation} 
where $f(\omega)=(e^{\omega/T}+1)^{-1}$ is the Fermi distribution
function. Hence we get
\begin{equation}
I=\frac{2e}{\hbar}Re\left[\sum_{{\bf pq}}|T_{{\bf
      pq}}|^2\int_{0}^{\infty}\frac{d\omega}{2\pi}\tanh\left(\frac{\omega}{2T}\right)
  \left(\mathcal{F}_{\uparrow\downarrow}^{R}({\bf
      q},\omega)\overline{\mathcal{F}}^{R}_{\downarrow\uparrow}({\bf
      p},\omega)-\mathcal{F}_{\uparrow\downarrow}^{A}({\bf
      q},\omega)\overline{\mathcal{F}}^{A}_{\downarrow\uparrow}({\bf
      p},\omega)\right)  \right].
\end{equation}

\begin{figure}[!htb]
\begin{center}
\includegraphics[scale=0.40,angle=0.1,keepaspectratio]{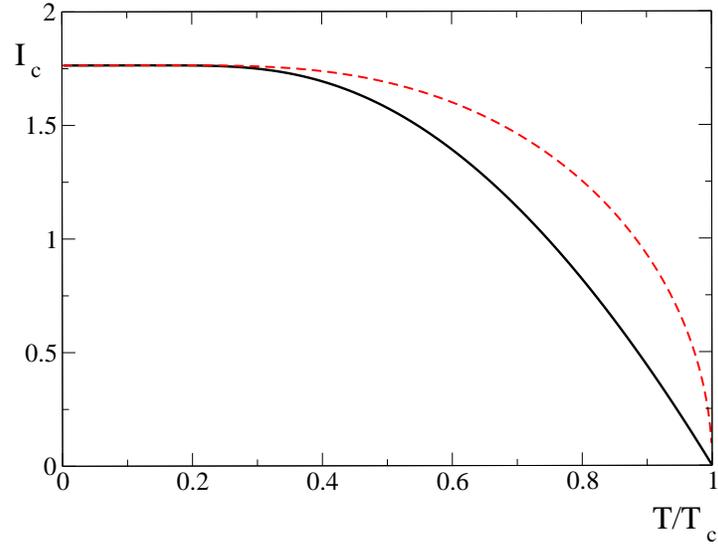}
\end{center}
\caption{\em Josephson critical current defined by Eq. \eqref{curr_fin} in the
  units of $\pi T_c/2eR_n$ (solid line). The dashed line shows the temperature
  dependence of the BCS gap $\Delta/T_c$.}
\label{current}
\end{figure}

Substituting the explicit expressions for the retarded and the advanced Gor'kov
functions \cite{AGD} we obtain
\begin{eqnarray}
I&=&\frac{2e}{\hbar}Re\Big[\sum_{{\bf pq}}|T_{{\bf
      pq}}|^2\int_{0}^{\infty}\frac{d\omega}{2\pi}\tanh\left(\frac{\omega}{2T}\right)\Big(\frac{-\Delta_{\bf
        q}}{(\omega+i\delta)^2-\epsilon_{\bf q}^2-|\Delta_{\bf q}|^2}\frac{-\Delta_{\bf
        p}^*}{(\omega+i\delta)^2-\epsilon_{\bf p}^2-|\Delta_{\bf p}|^2}
    \nonumber \\
&-&\frac{-\Delta_{\bf
        q}}{(\omega-i\delta)^2-\epsilon_{\bf q}^2-|\Delta_{\bf q}|^2}\frac{-\Delta_{\bf
        p}^*}{(\omega-i\delta)^2-\epsilon_{\bf p}^2-|\Delta_{\bf p}|^2}\Big) \Big].
\end{eqnarray}
We assume that superconducting gaps in the left and right leads are
momentum-independent $\Delta_{\bf q}\equiv \Delta_1$ and $\Delta_{\bf p}\equiv
\Delta_2$, and their product gives $\Delta_1 \Delta_2^*=|\Delta_1||\Delta_2|e^{i\phi}$, where $\phi$
is the phase difference between two superconductors. We also take into account
that the integrand is purely imaginary, so that the current becomes
\begin{eqnarray} 
I&=&\frac{2e}{\hbar}|\Delta_1||\Delta_2|(i\sin(\phi))\Big[\sum_{{\bf pq}}|T_{{\bf
      pq}}|^2\int_{0}^{\infty}\frac{d\omega}{2\pi}\Big(\frac{1}{(\omega+i\delta)^2-\epsilon_{\bf q}^2-|\Delta_{1}|^2}\frac{1}{(\omega+i\delta)^2-\epsilon_{\bf p}^2-|\Delta_{2}|^2}
    \nonumber \\
&-&\frac{1}{(\omega-i\delta)^2-\epsilon_{\bf q}^2-|\Delta_{1}|^2}\frac{1}{(\omega-i\delta)^2-\epsilon_{\bf p}^2-|\Delta_{2}|^2}\Big)\tanh\left(\frac{\omega}{2T}\right) \Big].
\end{eqnarray}
We thus derived microscopically that the current between two superconductors
is proportional to $\sin \phi$
\begin{equation}
I=I_c\sin \phi,
\end{equation}
where $I_c$ is the critical supercurrent
\begin{eqnarray}
I_c&=&\frac{2ie}{\hbar}|\Delta_1||\Delta_2|\sum_{{\bf pq}}|T_{{\bf
      pq}}|^2\int_{0}^{\infty}\frac{d\omega}{2\pi}\Big(\frac{1}{(\omega+i\delta)^2-\epsilon_{\bf q}^2-|\Delta_{1}|^2}\frac{1}{(\omega+i\delta)^2-\epsilon_{\bf p}^2-|\Delta_{2}|^2}
    \nonumber \\
&-&\frac{1}{(\omega-i\delta)^2-\epsilon_{\bf q}^2-|\Delta_{1}|^2}\frac{1}{(\omega-i\delta)^2-\epsilon_{\bf p}^2-|\Delta_{2}|^2}\Big)\tanh\left(\frac{\omega}{2T}\right).
\end{eqnarray}
Using standard rules of contour integration we can replace the integral over
$\omega$ by a sum over discrete Matsubara frequencies $\omega_n=(2n+1)\pi
T$ \cite{Mahan}
\begin{equation}
I_c=\frac{4e}{\hbar}|\Delta_1||\Delta_2|T\sum_{\bf pq}|T_{{\bf
      pq}}|^2\sum_{n=-\infty}^{\infty}\frac{1}{\omega_n^2+\epsilon_{\bf q}^2+|\Delta_{1}|^2}\frac{1}{\omega_n^2+\epsilon_{\bf p}^2+|\Delta_{2}|^2}.
\end{equation}
The summation over ${\bf p}$ and ${\bf q}$ can be replaced by an integral,
which can be taken
\begin{equation}
I_c=\frac{4\pi^2e}{\hbar}|\Delta_1||\Delta_2|N_1(0)N_2(0)\;|T_0|^2\;T\sum_{n=-\infty}^{\infty}\frac{1}{\sqrt{\omega_n^2+\Delta_1^2}}\frac{1}{\sqrt{\omega_n^2+\Delta_2^2}},
\end{equation}
where $N_1(0)$ and $N_2(0)$ are the densities of states at the Fermi energy in
the normal state of left and right lead correspondingly.
When the two gaps are equal to each other $\Delta_1=\Delta_2\equiv \Delta$, this expression
takes a simple form
\begin{equation}
I_c=\frac{2\pi^2e}{\hbar}\Delta\; N_1(0)N_2(0)\;|T_0|^2 \tanh\left(\frac{\Delta}{2T} \right).
\end{equation}
One usually introduces the so-called resistance of the tunneling junction in
the normal state $R_n$
\begin{equation}
\frac{1}{R_n}=4\pi e^2N_1(0)N_2(0)\;|T_0|^2, 
\end{equation}
so that
\begin{equation}
I_cR_n=\frac{\pi \Delta}{2e}\tanh\left(\frac{\Delta}{2T} \right).
\label{curr_fin}
\end{equation}
In Fig. \ref{current} we depict the temperature behavior of the Josephson
critical current $I_c$, normalized by $\pi T_c/2eR_n$. In order to obtain this
dependence we had to solve the standard BCS gap equation, and we present the
temperature dependence of the gap in the same Fig. One can see, that the critical current is
monotonously decreasing with temperature rather similar to the gap
behavior. Near $T_c$ the current is proportional to $\Delta^2$ and is
therefore linear in $(T_c-T)$.

At zero temperature the critical supercurrent for superconductors
with different gaps is 
\begin{equation}
I_c^0=\frac{2\Delta_1(0) \Delta_2(0)}{eR [\Delta_1(0)+\Delta_2(0)]}K\left(\frac{|\Delta_1(0)-\Delta_2(0)|}{\Delta_1(0)+\Delta_2(0)}\right),
\end{equation}
where $K(x)$ is a complete elliptic integral of the first kind. In case of
$\Delta_1(0)=\Delta_2(0)\equiv \Delta_0$ we get a simple expression 
\begin{equation}
I_c^0R_n=\frac{\pi}{2e} \Delta_0.
\end{equation}

\vspace{0.5cm}

\chapter{Josephson effect in Bose-Einstein condensates}

\vspace{0.5cm}

\begin{figure}[!htb]
\begin{center}
\includegraphics[scale=0.5,angle=0.1,keepaspectratio]{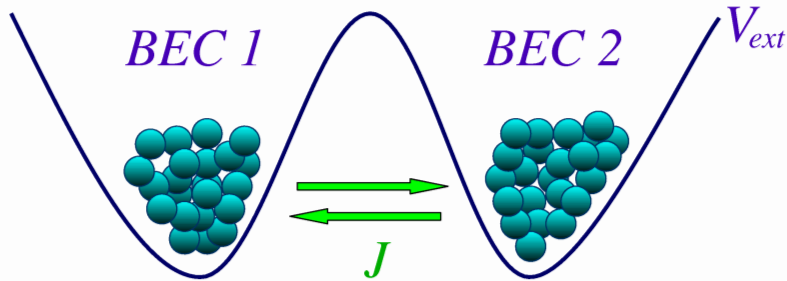}
\end{center}
\caption{\em Bose Josephson junction: two Bose-Einstein condensates are
  trapped in an external double-well potential $V_{ext}$.}
\label{doublewell}
\end{figure}

Condensates of cold alkali atoms \cite{Dalfovo,Leggett} provide a unique opportunity
to realize and to control
Josephson effect in a {\it weakly} interacting bosonic system. It was first
predicted in \cite{Juha}. A weak link
between two condensates can be realized in a double-well external potential
$V_{ext}$ (Fig. \ref{doublewell}). 
An interacting system of bosons confined in such a  potential is described by a general Hamiltonian
\begin{eqnarray}
H&=&\int d{\bf r}\;\hat \Psi^{\dagger}({\bf
  r},t)\left[-\frac{\hbar^2}{2m}\triangledown^2+V_{ext}({\bf r})\right]\hat\Psi({\bf
  r},t) \nonumber \\ &+&\frac{1}{2}\int d{\bf r}\int d{\bf r'}\hat\Psi^{\dagger}({\bf r},t)
   \hat\Psi^{\dagger}({\bf r'},t)V({\bf r}-{\bf r'})
   \hat\Psi({\bf r'},t) \hat\Psi({\bf r},t).
\end{eqnarray}
Here $\hat\Psi({\bf r},t)$ is the bosonic field operator and $V({\bf r}-{\bf r'})$ is
a two-particle interaction.

At low temperatures an experimentally realized gas of bosons is very dilute and particles are weakly
interacting, one can therefore introduce a contact interaction between the
particles
\begin{equation}
V({\bf r}-{\bf r'})=g\delta({\bf r}-{\bf r'}),
\end{equation}
where $g=4\pi \hbar^2 a_s/m$ with $a_s$ being the $s-$wave scattering
length. According to the Bogoliubov approximation one can consider the bosonic
field operator as a sum of a classical field (condensate wave function,
representing the condensate order parameter) and
excitations
\begin{equation}
\hat\Psi({\bf r},t)=\Psi({\bf r},t)+\delta \hat \Psi({\bf r},t). 
\end{equation} 
In the mean-field description we can neglect the excitations due the
smallness of the interaction term, so that our Hamiltonian becomes essentially
classical
\begin{equation}
H=\int d{\bf r}\;\Psi^{\dagger}({\bf
  r},t)\left[-\frac{\hbar^2}{2m}\triangledown^2+V_{ext}({\bf r})\right]\Psi({\bf
  r},t) +\frac{g}{2}\int d{\bf r}\Psi^{\dagger}({\bf r},t)
   \Psi^{\dagger}({\bf r},t)\Psi({\bf r},t) \Psi({\bf r},t).
\label{cl}
\end{equation}
For the Josephson effect to occur only a small overlap of the condensate wave-functions
is sufficient, and we can assume that the condensate wave-function $\Psi({\bf
  r},t)$ is given by the sum
of the order parameters for each well \cite{Smerzi97}
\begin{equation}
\Psi({\bf r},t)=\Psi_1({\bf r},t)+\Psi_2({\bf r},t)=\varphi_1({\bf
  r})\psi_1(t)+\varphi_2({\bf r})\psi_2(t).
\end{equation}
Here $\varphi_1({\bf r})$ and $\varphi_2({\bf r})$ are the ground state
solutions for isolated traps \cite{Smerzi97,Milburn}, and  
\begin{equation}
\psi_i(t)=N_i(t)e^{i\phi_i(t)}
\end{equation}
is  the complex condensate order parameter with $N_i$ being the number
of particles in the $i$-th well, and $\phi_i$ is the phase of the
condensate in the same well. With these notations the Hamilton function \eqref{cl}
takes the form
\begin{equation}
H=E_1N_1+E_2N_2+\frac{U_1}{2}N_1^2+\frac{U_2}{2}N_2^2+2J\sqrt{N_1N_2}\cos(\phi),
\label{ham_1}
\end{equation}
where $\phi$ is the phase difference between the wells,
\begin{eqnarray}
U_i&=&g\int d{\bf r}|\varphi_i|^4, \\
E_i&=&\int d{\bf r}\left(\frac{\hbar^2}{2m}|\triangledown
  \varphi_i({\bf r})|^2+\varphi_i^2({\bf r})V_{ext}({\bf r})   \right),
\end{eqnarray}
and $J$ is the Josephson coupling
\begin{equation}
J=-\int d{\bf r}\left(\frac{\hbar^2}{2m}\triangledown
  \varphi_1({\bf r})\triangledown
  \varphi_2({\bf r})+\varphi_1({\bf r})\varphi_2({\bf r})V_{ext}({\bf r})  \right).
\end{equation}
One can reexpress the Hamiltonian \eqref{ham_1} in terms of a particle
imbalance $n$ 
\begin{equation}
n=\frac{N_1-N_2}{N_1+N_2}
\end{equation}
so that the effective (dimensionless) Josephson Hamiltonian reads
\begin{equation}
H=\frac{\Lambda}{2}n^2-\sqrt{1-n^2}\cos \phi+\Delta E\; n.
\label{can_ham}
\end{equation}
We see that in this case only two parameters determine the behavior of the system: the
effective interaction 
\begin{equation}
\Lambda=(U_1+U_2)(N_1+N_2)/(4J)
\end{equation}
and the effective chemical potential difference
\begin{equation}
\Delta E=\frac{E_1-E_2}{2J}+\frac{(U_1-U_2)(N_1+N_2)}{4J}. 
\end{equation}

\begin{figure}[!htb]
\begin{center}
\includegraphics[scale=0.55,angle=0.1,keepaspectratio]{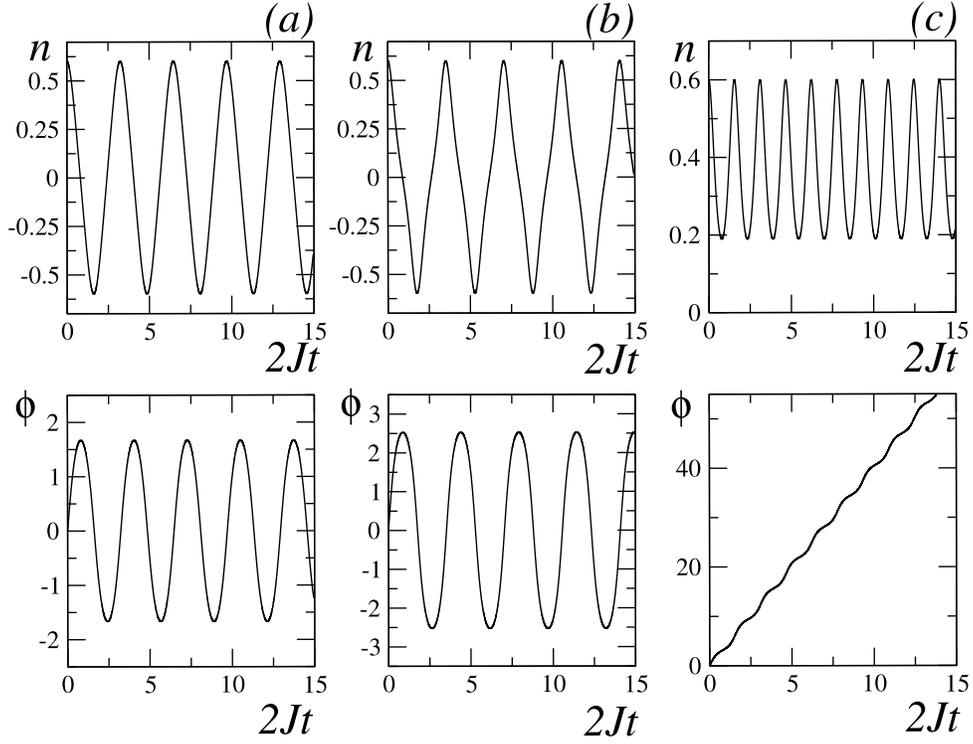}
\end{center}
\caption{\em Temporal oscillations of the particle imbalance $n$ and phase difference
  $\phi$ for $\Lambda=5$ (a), $\Lambda=9$ (b), and $\Lambda=11$
  (c). Initially $n(0)=0.6$ and $\phi(0)=0$.}
\label{imbalance}
\end{figure}

\begin{figure}[!htb]
\begin{center}
\includegraphics[scale=0.35,angle=0.1,keepaspectratio]{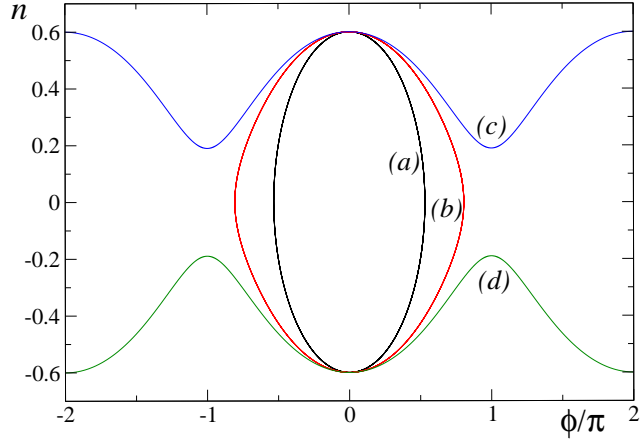}
\end{center}
\caption{\em Phase space plot: (a), (b) and (c) for same parameters as in
  Fig.\ref{imbalance}, (d) is for $\Lambda=11$, $n(0)=-0.6$, $\phi(0)=2\pi$. }
\label{phasespace}
\end{figure}

As the particle number operator $\hat n$ and phase operator $\hat \phi$ are canonically
conjugated variables, in the classical case one can identify a corresponding
Poisson bracket with their commutator. We can then  derive the corresponding  equations of motion for the particle imbalance and phase difference
\begin{equation}
\dot n=-\frac{\partial H}{\partial \phi}, \quad \dot \phi=\frac{\partial
H}{\partial n}.
\end{equation} 
As a result we get
\begin{eqnarray}
\dot n&=&-\sqrt{1-n^2}\sin \phi, \\
\dot \phi&=&\Lambda n+\Delta E+\frac{n}{\sqrt{1-n^2}}\cos \phi.
\label{bjj_eqs}  
\end{eqnarray}
These equations can be solved exactly in terms of the elliptic functions
\cite{Raghavan99}. In Fig. \ref{imbalance} we show the numerical solutions of the
equations \eqref{bjj_eqs} for $\Delta E=0$. One can observe a qualitative
change in the oscillations after $\Lambda$ exceeds a certain ``crossover''
value $\Lambda_c$. For initial conditions as in Fig. \ref{imbalance} $\Lambda_c\sim
10$. For $\Lambda <\Lambda_c$ the time-average of the particle imbalance is
zero: $\langle n \rangle_t=0$. For larger values of $\Lambda$ the particle
imbalance oscillates around a finite value (in case of Fig. \ref{imbalance}
$\langle n \rangle_t=0.4$). It means that on average the
number of particles in one well is  larger than the number of particles
in the other well. This curious quantum phenomenon was termed {\it macroscopic
  quantum self-trapping} (MST) \cite{Smerzi97}.  This behavior can be also
seen in a phase portrait in Fig.\ref{phasespace}, showing the constant energy
trajectories for different $\Lambda$. The ``running'' trajectories (c) and (d) correspond to MST.

\begin{figure}[!htb]
\begin{center}
\includegraphics[scale=0.5,angle=0.1,keepaspectratio]{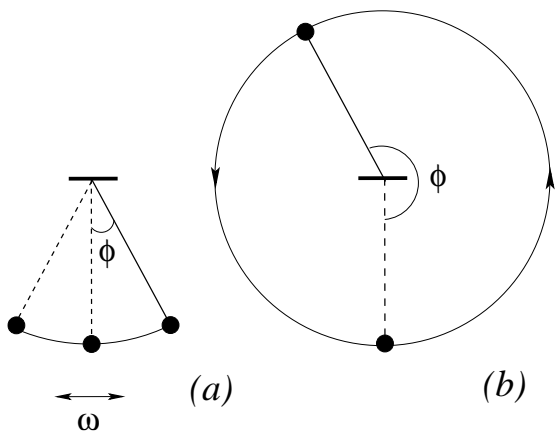}
\end{center}
\caption{\em Analogy with pendulum oscillations: MST state corresponds to the
  case (b), while standard Josephson oscillations of a bosonic junction to the case (a). }
\label{pendel}
\end{figure}

The occurrence of the MST phenomenon is readily understood if one remembers
that the canonical Josephson Hamiltonian 
\eqref{can_ham} in the small $n$ limit 
\begin{equation}
H=\frac{\Lambda}{2}n^2-\cos \phi+\Delta E\; n
\end{equation}
can be mapped onto a pendulum Hamiltonian with tilt angle $\phi$,
dimensionless angular momentum $p_{\phi}=n$, inverse mass $\Lambda$ and
applied torque $\Delta E$ \cite{Leggett}. For small $n$ the bosonic junction supports
small-amplitude Josephson ``plasma'' oscillations with the frequency
$\omega=\sqrt{\Lambda}/(2J)$ Fig. \ref{pendel}(a). A rotation of the pendulum in
Fig. \ref{pendel}(b) corresponds to the MST-state with a running phase. It is also
easy to understand the ``wiggles'' in the phase dynamics in Fig. \ref{imbalance}
(c), as the pendulum always slows down at its highest point. 

The physical behavior of the Bose Josephson junction described by the
Hamiltonian \eqref{can_ham} is more complicated due to the additional factor
$\sqrt{1-n^2}$. In the language of pendulum analogy it means that the length
of the pendulum is not rigid anymore, but varies with time. This leads to 
additional fixed points in comparison with the previous, simple example (for a
detailed description see \cite{Raghavan99,Shenoy02}).

The value of $\Lambda_c$ which determines the crossover to the MST state
is defined by the condition
\begin{equation}
H_0\equiv H(n(0),\phi(0))=\frac{\Lambda}{2}n(0)^2-\sqrt{1-n(0)^2}\cos(\phi(0))>1,
\end{equation}
so that
\begin{equation}
\Lambda_c=2\frac{1+\sqrt{1-n(0)^2}\cos(\phi(0))}{n(0)^2}.
\end{equation}
It means that $\Lambda_c$ can be relatively easy controlled in an experiment by varying
the initial conditions for two condensates. This property was used by
experimentalists and the predicted in \cite{Smerzi97} behavior of the Bose
Josephson junction was successfully observed experimentally \cite{Albiez05}. Josephson
oscillations of the particle imbalance with $\langle n \rangle_t=0$ were
observed for $n(0)\approx 0.28(6)$ and $\phi(0)\approx 0$. The MST regime
was achieved with $n(0)\approx 0.62(6)$ and $\phi(0)\approx 0$, the
parameter $\Lambda$ in both cases is estimated to have a fixed value of
$15(3)$. 

The analogue of a.c. and d.c. Josephson effect discussed for superconductors
in Section \ref{sc} has been  recently observed in a Bose Josephson
junction \cite{Levy}.

\vspace{0.5cm}

\chapter{Josephson effect in superfluid He}

\vspace{0.5cm}

Since superfluid Helium possesses phase rigidity, one would expect the Josephson
effect to occur between two weakly coupled Helium systems. Although liquid
Helium was discovered more than seventy years
ago (in 1937 by P. Kapitsa, J. F. Allen and D. Misener), it took a long time
before the Josephson effect has been finally observed: in 1997 in superfluid
$^3$He \cite{Pereverzev} and in 2001 in $^4$He \cite{Sukhatme}. The main
obstacle for the observation of the effect is a very
small coherence (healing) length of Helium: $50$ nm for $^3$He and even
smaller, of the order of  $0.1$ nm for $^4$He. It took thus almost 60 years to
overcome two main technical difficulties: (i) the creation of the weak link itself
- a structure with small apertures with dimensions of the scale of the
coherence length, (ii) measurement of tiny mass currents which would flow
through such a structure due to the Josephson effect. 

Note, that in case of liquid Helium, one can not apply an external
electromagnetic potential
difference to the system, as in the case of superconductors, neither  can one modify the
trapping potential to simulate this effect as in Bose-Einstein condensates of
cold atoms. For Helium the role of external potential $V$ is played by
pressure $P$, so that both  Josephson relations \eqref{eq1} and
\eqref{eq2} remain the same with chemical potential difference proportional to pressure:
\begin{equation}
\Delta \mu= \frac{\Delta P m}{\rho}.
\end{equation}
Here $m$ is the mass of either the $^4$He atom, or twice the  $^3$He atomic
mass ($^3$He is a fermionic system and its superfluidity is induced by
coupled fermions), $\rho$ is the liquid density. Applied pressure difference
will induce therefore an oscillating mass supercurrent with the frequency
$\omega_f=\Delta P m/\rho h$.

In the experiment \cite{Pereverzev} two $^3$He systems are separated by a
membrane with numerous apertures only 100 nm in diameter. The healing length at
the given experimental temperature was slightly exceeding the aperture
diameter. The great number of apertures (more than 4000) served to coherently increase the
monitored supercurrent, which was otherwise too tiny to be resolve in the
measurement. Another soft membrane controlled by an applied bias was used
in order to create an external pressure difference. Any displacements of the membrane due to the
supercurrent were monitored. Finally, the signals of the
supercurrent frequency obtained in this way were amplified and connected to
audio head-phones, and the listener could literally hear the effect of
coherent quantum oscillations between weakly coupled superfluids. It sounded
like a whistle smoothly drifting from high to low frequency while the pressure
relaxed to its zero value. The dependence of the supercurrent frequency on
$\Delta P$ has been found to be perfectly linear \cite{Pereverzev}. 

In $^4$He the regime of ordinary Josephson oscillations  across an aperture was for a
long time believed to be unobservable due to the strong fluctuations of the order
parameter in the volume with dimensions of the order of coherence length. In
the experiment, however, all the difficulties have been recently overcome and
a clear, unsmeared signature of Josephson oscillations has been found \cite{Sukhatme}. 

\chapter{Outlook}

\begin{figure}[!htb]
\begin{center}
\includegraphics[scale=0.5,angle=0.1,keepaspectratio]{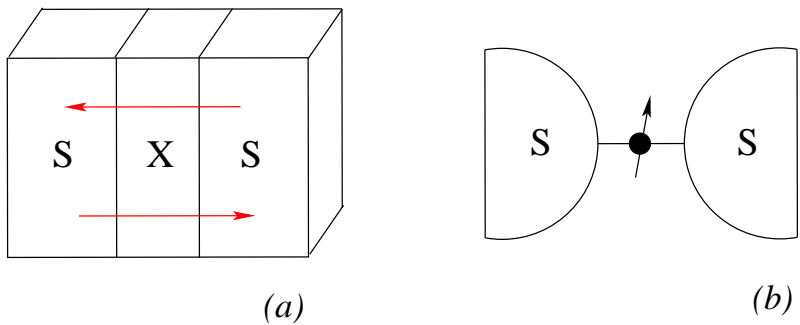}
\end{center}
\caption{\em Two examples of mesoscopic structures in which Josephson effect plays a role:  heterostructure with a normal or ferromagnetic layer
  embedded between two superconductors: X=N,FM (a); or a quantum dot
  coupled to superconducting leads (b).}
\label{hetero}
\end{figure}

The Josephson effect, predicted and discovered in the 1960-s in
superconducting systems, opened a broad
avenue of a new research area related to this phenomenon, which is still very
active. It is impossible to mention in a short review all the interesting
subfields which deal in one or another way with the Josephson phenomena. We
give therefore just a few examples. 

In addition to the contributions due to Cooper pair
tunneling discussed in Section \ref{sc}, one should take into account quasi-particle terms
\cite{Barone}, which we did not consider. Those contributions are especially important for the nonstationary Josephson
effect, i.e. in case of  the finite voltage applied to the junction. 
Many other effects influence the behavior of the particle tunneling between
two superconductors: various impurities, magnetic fields,
inhomogeneities, different pairing symmetries \cite{Barone,Kulik}.  

Many interesting phenomena arise due to the so-called proximity effect \cite{DeGennes}:
superconductivity penetrates up to a certain length scale into the neighboring
normal, or ferromagnetic material (X=N,FM as in Fig. \ref{hetero}(a)). In the
case of S-X-S heterostructure, shown in Fig. \ref{hetero}(a) the layer between
two superconductors does not need to be very thin for a supercurrent to occur
(for reviews see \cite{DeGennes,Golubov,Buzdin,Bergeret}). Due to the new experimental discoveries in 
superfluid Helium \cite{Pereverzev,Sukhatme}, the question arises whether an 
equivalent of an S-N-S structure can be created also in these systems. 

Another interesting system is a quantum dot coupled to two superconducting
leads shown in Fig. \ref{hetero}(b) (for review on the transport through
quantum dots see for instance \cite{Beenakker}). One can consider a similar
arrangement in a bosonic system \cite{Fischer}. 

Finally, a fundamental physical problem is a nonequilibrium Josephson effect. 
For example, in a bosonic system it is easily realized, as the barrier between the two
wells confining  condensates is ramped up in a nonadiabatic way
\cite{Albiez05}. This gives rise to quasi-particle excitations out of the
condensate \cite{Zapata,Mauro}. One can then develop a description in terms of
the Keldysh Green's functions \cite{Mauro}, which we mentioned  in the context
of the  superconducting Josephson junction in Section \ref{sc}.


\begin{thebibliography}{98}

\bibitem{Josephson} B. D. Josephson, Phys. Lett. {\bf 1}, 251 (1962). 

\bibitem{Barone} A. Barone, G. Paterno, {\it Physics and Applications of the
    Josephson Effect}, John Wiley and Sons, New York (1982). 

\bibitem{Anderson59} P. W. Anderson, J. Chem. Phys. Solids {\bf 11}, 26 (1959).

\bibitem{Abrikosov} A. A. Abrikosov, {\it Fundamentals of the Theory of Metals}, North
  Holland, Amsterdam, Oxford, New York, Tokio, 1988. 

\bibitem{Feynman} R. P.  Feynman, R. B. Leighton, and M. Sands. The
  Schr\"odinger equation in a classical context: A seminar on
  superconductivity. In {\it The Feynman Lectures on Physics}, Vol. III,
  Addison-Wesley (1965). 

\bibitem{Anderson} P. W. Anderson, Ravello-Lectures on the Many-Body
  Problem. 2, Ed. E. R. Gaianello, Acad. Press, p. 115 (1963).

\bibitem{Ambegaokar} V. Ambegaokar, and A. Baratoff, Phys. Rev. Lett. {\bf
    10}, 486 (1963), {\it ibid} {\bf 11}, 104 (1963).


\bibitem{Kulik} I. O. Kulik and I. K. Yanson, {\it The Josephson effect in
    superconductive tunneling structures}, Jerusalem 1972, translation from the
  book {\it Effekt DZhozefsona v sverkhprovodyashchikh tunnel'nykh
    strukturakh}, Nauka, Moskva 1970.

\bibitem{Mahan} Gerald D. Mahan, {\it Many-Particle Physics}, Plenum Press,
  New York and London, 1990.

\bibitem{Keldysh} L. V. Keldysh, Zh. Eksp. Teor. Fiz. {\bf 47}, 1515 (1964)
  [Sov. Phys.--JETP {\bf 20}, 1018 (1965)].

\bibitem{Rammer} J. Rammer and H. Smith, Rev. Mod. Phys. {\bf 58}, 323
  (1986).

\bibitem{Larkin75} A. I. Larkin, and Yu. N. Ovchninnikov,
  Zh. Eksp. Teor. Fiz. {\bf 68}, 1915 (1975) [Sov. Phys. - JETP {\bf 41}, 960 (1975)]. 

\bibitem{AGD} A. A. Abrikosov, L. P. Gor'kov, I. Ye. Dzyaloshinskii, {\it
    Quantum Field Theoretical Methods in Statistical Physics}, Pergamon, (1965).

\bibitem{Dalfovo} F. Dalfovo, S. Giorgini, L. P. Pitaevskii, S. Stringari,
  Rev. Mod. Phys. {\bf 71}, 463 (1999).

\bibitem{Leggett} A. J. Leggett, Rev. Mod. Phys. {\bf 73}, 307 (2001). 

\bibitem{Juha} J. Javanainen, Phys. Rev. Lett. {\bf 57}, 3164 (1986). 

\bibitem{Smerzi97} A. Smerzi, S. Fantoni, S. Giovanazzi, and S. R. Shenoy,
  Pys. Rev. Lett. {\bf 79}, 4950 (1997).

\bibitem{Milburn} G. J. Milburn, J. Corney, E. M. Wright, and D. F. Walls,
  Phys. Rev. A {\bf 55}, 4318 (1997).


\bibitem{Shenoy02} Subodh R. Shenoy, Pramana  {\bf 58}, 385 (2002).

\bibitem{Raghavan99} S. Raghavan, A. Smerzi, S. Fantoni, and S. Shenoy,
  Phys. Rev. A {\bf 59}, 620 (1999)

\bibitem{Albiez05} M. Albiez, R. Gati, J. F\"olling, S. Hunsmann,
  M. Cristiani, and M. K. Oberthaler, Phys. Rev. Lett. {\bf 95}, 010402 (2005).

\bibitem{Levy} S. Levy, E. Lahoud, I. Shomroni, and J. Steinhauer, Nature {\bf
  449}, 579 (2007).

\bibitem{Pereverzev} S. V. Pereverzev, A. Loshak, S. Backhaus, J. C. Davis,
  and R. E. Packard, Nature {\bf 388}, 449 (1997).

\bibitem{Sukhatme} K. Sukhatme, Y. Mukharsky, T. Chui, and D. Pearson, Nature
  {\bf 411}, 280 (2001).

\bibitem{DeGennes} P.G. de Gennes, Rev. Mod. Phys. {\bf 36}, 225 (1964).

\bibitem{Golubov} A.A. Golubov, M. Yu. Kupriyanov, E. Il'ichev,
Rev. Mod. Phys. {\bf 76}, 411 (2004).

\bibitem{Buzdin} A. Buzdin, Rev. Mod. Phys. {\bf 77}, 935 (2005). 

\bibitem{Bergeret} F. S. Bergeret, A. F. Volkov, and K. B. Efetov,
  Rev. Mod. Phys. {\bf 77}, 1321 (2005). 

\bibitem{Beenakker} C. W. J. Beenakker, and H. Van Houten, Solid State Physics
  {\bf 44}, 1 (1991).

\bibitem{Fischer} U. R. Fischer, C. Iniotakis, and A. Posazhennikova,
  Phys. Rev. A {\bf 77}, 031602 (R) (2008).

\bibitem{Zapata} I. Zapata, F. Sols, and A. J. Leggett, Phys. Rev. A {\bf 67},
  021603(R) (2003).

\bibitem{Mauro} M. Trujillo Martinez, A. Posazhennikova, and J. Kroha,
  {\it cond-mat preprint} 0903.5459 (2009). 





\end{thebibliography}
\end{document}